\documentclass[aps,prd,showpacs,twocolumn,superscriptaddress,nofootinbib,preprintnumbers]{revtex4-1} 

\makeatletter
\def\p@subsection{}
\makeatother

\usepackage{color,graphicx,amsmath,amssymb,lipsum}
\usepackage[colorlinks=true,citecolor=blue,linkcolor=blue,urlcolor=blue, backref=false,pdfborder={0 0 0}]{hyperref}

\usepackage[normalem]{ulem}
\usepackage[utf8]{inputenc} 
\usepackage{mathtools}

\usepackage{float}

\usepackage{ulem}
\usepackage{diagbox}

\usepackage[dvipsnames]{xcolor}
\usepackage{mathrsfs}

\def\le{\left(}
\def\ri{\right)}

\newcommand{\be}{\begin{equation}}
\newcommand{\ee}{\end{equation}}
\newcommand{\beqa}{\begin{eqnarray}}
\newcommand{\eeqa}{\end{eqnarray}}

\renewcommand\r{\rho}

\def\d{\partial}
\newcommand{\bseq}{\begin{subequations}}
\newcommand{\eseq}{\end{subequations}}

\def\gsim{\raise0.3ex\hbox{$\;>$\kern-0.75em\raise-1.1ex\hbox{$\sim\;$}}}
\def\lsim{\raise0.3ex\hbox{$\;<$\kern-0.75em\raise-1.1ex\hbox{$\sim\;$}}}

\def\beqn#1{\begin{equation}\label{#1}}
\def\eeqn{\end{equation}}

\def\beqa#1{\begin{eqnarray}\label{#1}}
\def\eeqa{\end{eqnarray}}

\def\Z2{$\mathcal{Z_2}$}

\newcommand {\ignore}[1]{}

\begin{document}

\preprint{INR-TH-2021-003}

\title{Hidden Symmetry of Vanishing Love}

\author{Panagiotis Charalambous}
\email{pc2560@nyu.edu}
\affiliation{Center for Cosmology and Particle Physics, Department of Physics, New York University, New York, NY 10003, USA}
\author{Sergei Dubovsky}
\email{sergei.dubovsky@gmail.com}
\affiliation{Center for Cosmology and Particle Physics, Department of Physics, New York University, New York, NY 10003, USA}
\author{Mikhail M. Ivanov}
\email{mi1271@nyu.edu}
\affiliation{Center for Cosmology and Particle Physics, Department of Physics, New York University, New York, NY 10003, USA}
\affiliation{Institute for Nuclear Research of the
Russian Academy of Sciences, \\ 
60th October Anniversary Prospect, 7a, 117312
Moscow, Russia
}

\begin{abstract} 
We show that perturbations of massless fields in the Kerr black hole background
enjoy a hidden $SL(2,\mathbb{R})\times {U}(1)$ (``Love") symmetry in the properly defined near zone approximation. Love symmetry 
 mixes IR and UV modes.
Still, this approximate symmetry allows us to derive exact results  about static tidal responses. 
 Generators of the Love symmetry are globally well defined and have a smooth
Schwarzschild limit. 
Generic regular solutions of the near zone Teukolsky equation 
form infinite-dimensional $SL(2,\mathbb{R})$ representations. In some special cases ($\hat{\ell}$ parameter is an integer), these  are highest weight representations.
This is the situation
that corresponds to vanishing Love numbers. 
In particular, static perturbations of 
 four-dimensional Schwarzschild black holes  belong to finite-dimensional representations. Other known facts about static Love numbers also acquire an elegant explanation in terms of the $SL(2,\mathbb{R})$ representation theory.
\end{abstract}

\maketitle

\section{Introduction}
The  LIGO  detection of gravitational waves \cite{Abbott:2016blz} from inspiralling black hole binaries opened an era of precision black hole physics.
The worldline effective theory~\cite{Goldberger:2004jt,Goldberger:2005cd,Porto:2016pyg} provides
 an efficient modern toolbox for analytical calculations of the waveforms from binary inspirals and for interpreting the results. In this framework each of the individual black holes in the binary is treated as a point-like particle. Finite size effects are captured by higher-dimensional operators on the worldline.
 This approach is analogous to the multipole expansion in electrodynamics.

 Wilson coefficients in front of  operators with a quadratic dependence on external fields are called Love numbers. They characterize black hole tidal responses~\cite{Binnington:2009bb}.
Remarkably, static Love numbers, which determine response to time-independent external fields, are found to vanish in four-dimensional
Einstein theory both for spherical and spinning black holes~\cite{Fang:2005qq,Damour:2009vw,Binnington:2009bb,Kol:2011vg,Hui:2020xxx,Chia:2020yla,Charalambous:2021mea}.
In this regard, black holes are called the most rigid objects in the Universe. 
 In the worldline effective field theory context, this implies that all quadratic finite-size operators without time derivatives vanish for black holes,
which represents an outstanding naturalness problem in the context of the worldline
effective  theory~\cite{Porto:2016zng}.

In four dimensions, static Love numbers vanish for 
perturbing fields of all spins and for an arbitrary multipolar
index $\ell$. To add to the puzzle, the situation is far more complicated  for higher-dimensional Schwarzschild black holes~\cite{Kol:2011vg,Hui:2020xxx}. Static Love numbers are nonzero in higher dimensions for generic multipolar indices $\ell$. However, they do vanish for some special values of $\ell$, and for some other special values they exhibit classical renormalization group running.

This  intricate pattern calls for a novel (``Love") symmetry of black holes which would account 
for the peculiar behavior of static Love numbers. 
In this Letter, we identify
such a  symmetry.  

\section{Near Zone Expansion}
We start with  the simplest case of a  massless scalar field $\varphi$  in the Kerr background. The resulting Klein--Gordon equation is known to be separable in the Boyer--Lindquist coordinates\footnote{Our conventions for the Kerr metric are summarized in  Appendix~\ref{App:met}.}. After writing 
\be
\label{sep}
\varphi=\Phi(t,r,\phi)S(\theta)=R(r)S(\theta)e^{-i\omega t+im\phi}
\ee
one arrives at the spin weight $s=0$ Teukolsky equation  \cite{Teukolsky:1973ha} for the radial function,
\be
\label{Teukolsky}
\d_r\le \Delta \d_r R\ri+\le V_0+\epsilon V_1
\ri R=\ell(\ell+1)R\;,
\ee
where
\begin{gather}
\label{V0}
V_0={(2Mr_+)^2\over \Delta}\le{(\omega -\Omega m)^2}-4\omega\Omega m  {r-r_+\over r_+-r_-}\ri\;,\\
\label{V1}
 V_1=\frac{2M(\omega a m\beta +4M^2\omega^2
 r_+
 )}{r_+ (r-r_-)}
 +\omega^2(r^2+2Mr+4M^2)\,,
\end{gather}
and 
we have introduced 
\be
 \beta= \frac{4M r_+}{r_+-r_-}\;,
\ee
and $\ell(\ell+1)$ is the eigenvalue of the angular operator (\ref{ang}), while $\Omega=a/2Mr_+$ is black hole's angular velocity. Note that, in general, $\ell$ is not an integer.
Here, $\epsilon$ is a formal parameter of the near zone expansion. 
For the physical Kerr background $\epsilon=1$, while throughout this Letter we are working in the leading near zone approximation, $\epsilon=0$. As follows from (\ref{V1}), the leading near zone approximation is accurate provided 
\be
\label{eq:NZ}
\omega r \ll 1\,,\quad~M\omega \ll 1\,.
\ee
The range of validity of the near zone approximation covers the near horizon region $r\gtrsim r_+$ and overlaps with the asymptotically flat region $r\gg r_+$. 

It is important to note that the near zone expansion is different from the low frequency expansion because one keeps some frequency dependent terms in  the Teukolsky equation even at the leading order in the near zone expansion. Nevertheless, it provides an accurate approximation at low frequencies. 
In particular, the leading near zone approximation produces exact answers for $\omega=0$ quantities, such as  static tidal responses. 

Related to this, there is an ambiguity in how one defines the near zone expansion associated with a freedom to move $\omega$ dependent terms between $V_0$  and $V_1$  as soon as  $V_1$ stays  finite at the horizon. Other choices of the near zone split can be found in, e.g., ~\cite{1973JETP...37...28S,Maldacena:1997ih,Castro:2010fd}.
 
 \section{Love Symmetry}
 The reason for our choice is related to the following crucial observation.
Let us consider three vector fields of the form
\be
\label{eq:vec0}
\begin{split}
& L_0 = - \beta\d_t\,,\\
& L_{\pm 1} =  e^{\pm\beta^{-1} t}\le
\mp\Delta^{1/2}\d_r + \beta\d_r(\Delta^{1/2})\d_t +
\frac{a}{\Delta^{1/2}}
\d_\phi
\ri\,.
\end{split}
\ee
It is straightforward to check that these fields satisfy the $SL(2,\mathbb{R})$ algebra,
\be
\label{eq:walg}
[L_n, L_m]=(n-m)L_{n+m}\,, \quad n,m=-1,0,1\,.
\ee
Using the quadratic Casimir of this algebra
\be
\begin{split}
\mathcal{C}_2
&\equiv L_0^2 -\frac{1}{2}(L_{-1}L_{1}+L_{1}L_{-1})\\
\end{split}
\ee
one finds that the $\epsilon=0$ Teukolsky equation can be written as  
\be
\label{eq:KGeq}
\mathcal{C}_2\Phi= \ell(\ell+1)\Phi  \,.
\ee
Eigenvalues of the operator $L_0$
are given by 
\be
L_0 \Phi = i\beta\omega \Phi\equiv h\Phi\,. 
\ee
By  transforming into advanced or retarded coordinates it is straightforward to check that all three $SL(2,\mathbb{R})$ generators are regular at the black hole horizon. As a result, regular solutions of the near zone Teukolsky equation form $SL(2,\mathbb{R})$ representations even though the symmetry is ``hidden"---it does not correspond to an isometry of the background. We will refer to this hidden symmetry as the Love symmetry.

The above properties of the Love symmetry can be contrasted 
with the noncritical Kerr/CFT proposal~\cite{Castro:2010fd}. It was observed there that,  for a different choice of the near zone split,
 the leading order Teukolsky equation enjoys a local hidden 
$SL(2,\mathbb{R})_L\times SL(2,\mathbb{R})_R$ conformal symmetry. However, the 
corresponding vector fields are not well defined globally, because they do not respect the $\phi\to\phi+2\pi$ periodicity. As a result, regular solutions of the Teukolsky equation do not form $SL(2,\mathbb{R})_L\times SL(2,\mathbb{R})_R$ representations.

Furthermore,  the Love symmetry generators (\ref{eq:vec0}) have a smooth Schwarzschild limit, which is not the case for the Kerr/CFT $SL(2,\mathbb{R})_L\times SL(2,\mathbb{R})_R$. At $a=0$ vector fields (\ref{eq:vec0}) reduce to the ones derived previously in~\cite{Bertini:2011ga}. 

These considerations suggest that the Love symmetry (\ref{eq:vec0}) may be a better starting point for a holographic description of Kerr black holes. This expectation is further supported by the observation that the $SL(2,\mathbb{R})\times U(1)$ symmetry which we found (where the $U(1)$ factor corresponds to axial rotations) matches the near horizon isometry of the extreme Kerr solution \cite{Bardeen:1999px,Guica:2008mu}.  A nonextreme Kerr black hole may be considered as an excitation above the leading Regge trajectory populated by extreme Kerr states\footnote{Recently, an analogous approach proved to be very useful for understanding the spectrum of Yang--Mills glueballs \cite{Dubovsky:2016cog}.}. From this viewpoint, it is natural to identify the hidden Love symmetry  (\ref{eq:vec0})  with the $SL(2,\mathbb{R})$ isometry of the extreme near horizon region. In an excited nonextreme state this symmetry gets spontaneously broken and, thus, ceases to be an isometry.

\section{Highest Weight Banishes Love}

To illustrate the power of the 
Love symmetry, let us apply it to explain properties of  static Love numbers. 
To define them, one 
looks at the large-$r$
behavior of the static radial 
solution
$R(r)$ with a fixed
growing asymptotics, determined by a source
at spatial 
infinity. 
Love numbers 
are defined as coefficients in front of decaying powers 
in this asymptotic expansion.
For black holes 
in four dimensional 
general relativity 
this radial solution
turns out to be a 
polynomial in $r$ (in the appropriate coordinates),
hence Love numbers
vanish identically.
We will show now that this polynomial form 
is dictated by the highest weight property 
of the corresponding 
$SL(2,\mathbb{R})$
representation.

Let us start with the Schwarzschild case, $a=0$. Generic solutions of the Teukolsky equation correspond to infinite-dimensional representations of the Love symmetry. However, $SL(2,\mathbb{R})$ algebra also has finite-dimensional representations for positive integer $\ell$ and integer $L_0$ eigenvalues $|h|\leq\ell$. These nonunitary representations can be obtained by a  ``Wick rotation" of the familiar unitary  $SO(3)$ angular momentum representations.

Furthermore, note that, at the leading order in the near zone expansion, the angular equation (\ref{ang}) turns into the standard equation for the associated Legendre polynomials,
and hence, $\ell$ is a positive integer\footnote{Of course, in the Schwarzschild case this is true without taking the near zone limit.}, satisfying $\ell\ge|m|$. 
This suggests that static Schwarzschild perturbations, which have $h=0$, belong to a finite-dimensional representation of the Love symmetry.
To prove this, let us consider the $h=-\ell$ highest weight vector $v_{-\ell,0}$,
\be
\label{highestw}
L_1v_{-\ell,0}=0\;,\;\; L_0v_{-\ell,0}=-\ell v_{-\ell,0}\;.
\ee
By making use of (\ref{eq:vec0}), one finds 
\be
\label{vl}
v_{-\ell,0}=e^{\ell\beta^{-1} t}\Delta^{\ell/2}\;,
\ee
where we set $m=0$ without  loss of generality.
As a consequence of the $SL(2,\mathbb{R})$ commutation relations $v_{-\ell,0}$ solves the Teukolsky equation (\ref{eq:KGeq}).
By transforming into advanced or retarded coordinates, one finds that this solution is regular at the black hole horizon.

As an aside, it is worth noting that conventionally one ensures regularity of the solutions to the Teukolsky equation by enforcing the incoming wave condition at the horizon \cite{1972ApJ...175..243P,Teukolsky:1973ha}. However, this criterion may fail for purely imaginary frequencies, which is the case for $v_{-\ell,0}$. In particular,
the lowest weight vector $\bar{v}_{\ell,0}$ satisfying 
\be
\label{highestw}
L_{-1}\bar{v}_{\ell,0}=0\;,\;\; L_0\bar{v}_{\ell,0}=\ell v_0\;,
\ee
takes form
\be
\label{vbl}
\bar{v}_{\ell,0}=e^{-\ell\beta^{-1} t}\Delta^{\ell/2}\;
\ee
and also provides us a regular at the horizon solution of (\ref{eq:KGeq}). Regularity of both (\ref{vl}) and (\ref{vbl}) is counterintuitive from the viewpoint of the incoming wave condition. Nevertheless, both solutions are regular as can be checked by transforming to the advanced or retarded coordinates.

One may obtain the rest of the representation by acting 
on the highest weight 
vector $v_{-\ell,0}$ 
with the lowering 
operator $L_{-1}$,
which increases $h$ by 
unity.
This way one arrives at 
the static solution with $h=0$ given by
\[
v_{-\ell,\ell}=L_{-1}^\ell v_{-\ell,0}\;.
\]
Since the highest weight 
vector $v_{-\ell,0}$ and $L_{-1}$ 
are both regular at the horizon, the same is true for $v_{-\ell,\ell}$
and all other states 
in the multiplet.
Now we can use the
$SL(2,\mathbb{R})$ algebra in the opposite 
direction.
We take
the static solution 
$v_{-\ell,\ell}$ and 
climb up to the 
highest weight state 
by applying $\ell$ times
the
raising operator
$L_{+1}$, i.e. $v_{-\ell,0}\propto L^\ell_{+1}v_{-\ell,\ell}$.
But the highest 
weight vector itself is annihilated by $L_{+1}$,
\be 
\label{eq:L1}
L_{+1}^{\ell+1}v_{-\ell,\ell}
\propto L_{+1}v_{-\ell,0}
=0\,.
\ee 
Additionally, it follows from (\ref{eq:vec0}) that
\be
\label{sphpol}
L_{+1}^{\ell+1}v(r)=(-1)^{\ell+1}e^{(\ell+1)\beta^{-1} t}\Delta^{{\ell+1}\over 2}\d_r^{\ell+1}v(r)\;,
\ee 
for any function $v(r)$ independent of $t$ and $\phi$. 
Then   
Eq.~\eqref{eq:L1} dictates
that the static solution $v_{-\ell,\ell}$ is an $\ell$-th degree polynomial in $r$. Given that the corresponding
Love number is defined as a coefficient in front of $r^{-\ell-1}$ in the $r\to \infty $ expansion of $v_{-\ell,\ell}$, we conclude that scalar static Love numbers of Schwarzchild black holes all vanish as a consequence of the $SL(2,\mathbb{R})$ algebra.
This result is exact
even though this derivation has been performed at the leading order in the near zone expansion.

Note that we could also have obtained a static regular solution with the same value of $\ell$ by acting with $L_1^{\ell}$ on the lowest weight vector
$\bar{v}_{\ell,0}$. From the uniqueness of the regular solution, it follows, then, that $\bar{v}_{\ell,0}$ belongs to the same $SL(2,\mathbb{R})$ representation,
\be
\bar{v}_{\ell,0}\propto L_{-1}^{2\ell}{v}_{-\ell,0} \equiv v_{-\ell,2\ell}\;.
\ee
This implies that the dimensionality of the corresponding representation is finite and is  equal to $2\ell+1$.

A large part of this argument proceeds unchanged for a rotating black hole. The first step is to look for the highest weight vector $v_{-\ell,0}(m)$.
Using (\ref{eq:vec0}) one finds
\be
\label{rothw}
v_{-\ell,0}(m)=e^{\ell\beta^{-1} t+im\phi}{(r-r_+)^{i{m\Omega\beta\over 2}+{\ell\over 2}}\over(r-r_-)^{i{m\Omega\beta\over 2}-{\ell\over 2}}}\;,
\ee
which is, again, regular at the horizon. Hence, the descendant vector $v_{-\ell,\ell}(m)$ is, again, a regular static solution annihilated by $L_1^{\ell+1}$.
The complication is that $v_{-\ell,\ell}(m)$ now depends on $\phi$, so a generalization of (\ref{sphpol}) is required. Inspecting the explicit expression for $L_1$ in (\ref{eq:vec0}) suggests the following ansatz for $v_{-\ell,\ell}(m)$,
\be
\label{ansatz}
v_{-\ell,\ell}(m)=e^{i m \phi}{\cal F}(r)v(r)\;,
\ee
where
\be
\label{ff}
{\cal F}(r)={(r-r_+)^{i{m\Omega\beta\over 2}}\over(r-r_-)^{i{m\Omega\beta\over 2}}}\;.
\ee
Indeed, then, one finds that 
\be
\label{rotpol}
L_1^{\ell+1}e^{i m \phi}
{\cal F}(r)v(r)
=(-1)^{\ell+1}e^{(\ell+1)\beta^{-1} t} 
{\cal F}(r)
\Delta^{\frac{\ell+1}{2}}
\d_r^{\ell+1}v(r)\;,
\ee
again implying that $v(r)$ is a degree $\ell$ polynomial in $r$. This result agrees with the brute force solution of the Teukolsky equation, which results in the explicit expression for $v(r)$ in terms of a hypergeometric function (see, e.g., \cite{Charalambous:2021mea}).

Naively, the expression (\ref{ansatz}) suggests the presence of a nontrivial tidal response associated with the  nonpolynomial form factor (\ref{ff}). However, 
as explained in \cite{Chia:2020yla,Charalambous:2021mea}, this response can be attributed to frame dragging.
It is purely dissipative and does not correspond to an effect of local worldline operators. The static Love numbers are still zero. An intuitive way 
to see this is to notice that the form factor (\ref{ff}) disappears completely if one were to perform a transform into the advanced coordinates. If we were to perform
the calculation in the advanced coordinates to start with, as was advocated in \cite{Poisson:2014gka}, the result would be purely polynomial.

Note that, unlike (\ref{vl}), the highest weight vector (\ref{rothw}) is regular only at the future (black hole) horizon. It is zero at the past (white hole) horizon and exhibits a
branch point singularity there. This is acceptable physically in the response calculations \cite{Teukolsky:1973ha}, because the white hole horizon of an eternal black hole is never 
present for physical black holes formed as a result of a collapse. This also clarifies a physical meaning of the prefactor (\ref{ff})---it signals the presence of a 
 singularity at the white hole horizon. In this case, static solutions belong to infinite-dimensional highest weight
$SL(2,\mathbb{R})$ representations (``Verma modules")---the lowest weight vector $\bar{v}_{\ell,0}(m)$ is singular at the future horizon at $\Omega m\neq 0$ (and regular at the white hole horizon) and, thus, belongs to a different representation.

To summarize, we see that the $SL(2,\mathbb{R})$ representation theory provides an elegant algebraic characterization for the properties of the static 
Love numbers. Vanishing Love numbers correspond to highest weight $SL(2,\mathbb{R})$ representations. In general, these are infinite dimensional, which corresponds to a singularity at the white hole horizon. Finite-dimensional representations (which necessarily exhibit highest and lowest weight properties simultaneously) arise when 
the corresponding solutions are regular at the horizon both in advanced and retarded coordinates.
 
 \section{Generalizations}
Remarkably, the puzzling properties of static Love numbers for higher-dimensional Schwarzschild black holes can also all be nicely phrased in terms of the representation theory. Spherical higher-dimensional black holes also exhibit a hidden  $SL(2,\mathbb{R})$ symmetry \cite{Bertini:2011ga}. Its generators are summarized 
in Appendix~\ref{sec:highd}. The near zone Teukolsky equation now takes the following form in $d$ spacetime dimensions, 
\be 
\begin{split}
\mathcal{C}_2\Phi = \hat{\ell}\left(\hat{\ell}+1\right) \Phi \,,\;\; \mbox{\rm with } \hat{\ell}={\ell\over d-3}\;.
\end{split}
\ee
For integer values of $\hat{\ell}$ one again arrives at finite-dimensional $SL(2,\mathbb{R})$ representations.
This is exactly the case when the static Love numbers were shown to vanish~\cite{Kol:2011vg,Hui:2020xxx}.

Generically, $\hat{\ell}$ is not an integer, the corresponding representations do not have the highest weight form and Love numbers do not vanish.
Still, the $SL(2,\mathbb{R})$ representation theory explains why these Love numbers do not exhibit logarithmic running. The point is that generically singular and regular solutions of the Teukolsky equation correspond to different $SL(2,\mathbb{R})$ representations. This provides a local criterion for selecting the regular one and
excludes the possibility of renormalization group running. 

This argument breaks down at half-integer $\hat{\ell}$'s. As $\hat{\ell}$ approaches a half-integer value, $SL(2,\mathbb{R})$ representations describing  regular and singular solutions become the same (see Chapter VII of~\cite{Vilenkin:2623281}). This makes it impossible to distinguish them locally and leads to a classical renormalization group running of Love numbers for half-integer $\hat{\ell}$'s~\cite{Kol:2011vg,Hui:2020xxx}. It appears that a proper analogy for this phenomenon is a resonance condition required for the logarithmic running to appear in conformal perturbation theory, c.f. \cite{zamolodchikov1989integrable,Konechny:2003yy}.

The arguments above can be straightforwardly extended to other bosonic fields in four dimensions. We provide the details in \cite{future} and present just a short summary here.
The generalization of the generators \eqref{eq:vec0}
for a generic massless field of spin weight $s$
is given by 
\be
\label{eq:vecs}
\begin{split}
 L^{(s)}_0 & = L_0 + s\,,\\
L^{(s)}_{\pm 1} & = 
L_{\pm 1}
-se^{\pm\beta^{-1} t}(1\pm 1)\d_r (\Delta^{1/2})\,.
\end{split}
\ee
The corresponding quadratic Casimir satisfies the spin weight $s$ Teukolsky
equation~\cite{Teukolsky:1972my,Teukolsky:1973ha} in the near zone approximation\footnote{As in the $s=0$ case discussed 
above, this near zone split is slightly 
different
from the one used in Refs.~\cite{Page:1976df,1974JETP...38....1S}.},
\begin{gather}
\mathcal{C}^{(s)}_2 \psi_s =  \le\mathcal{C}_2+
s(\d_r \Delta) \d_r 
+s \frac{2Mr_+(r_+-r_-)}{\Delta}\d_t \right.
\nonumber
\\
\left.
+s\frac{2(r-M)}{\Delta}a\d_\phi +s^2 +s
\ri\psi_s=
\ell(\ell+1)
\psi_s\,,
\end{gather}
where $\psi_0=\Phi$ is a test scalar field,  
$\psi_{\pm 1}$ are the Newman-Penrose-Maxwell scalars, from which one can extract the electromagnetic field
around the black hole~\cite{Teukolsky:1973ha},
and $\psi_{\pm 2}$ are the Newman-Penrose-Weyl scalars that can be used 
to reconstruct gravitational perturbations~\cite{Teukolsky:1973ha,Yunes:2005ve,LeTiec:2020bos}.
The structure of the symmetry algebra for a generic spin weight $s$ 
is identical to the 
scalar field case discussed above. Again, $\ell$ is an integer at the leading order in the near zone expansion, implying the highest weight property and the vanishing of all static Love numbers.

\section{Infinite Extension of Love}
Very general  arguments \cite{Hofman:2011zj} suggest that the $SL(2,\mathbb{R})\times {U}(1)$ symmetry discussed so far
is just a small part of a full infinite-dimensional algebra. Note that the proof of \cite{Hofman:2011zj} does not apply here directly, because it relies on unitarity, and the representations encountered above are all nonunitary. Nevertheless, there are indications that $SL(2,\mathbb{R})\times {U}(1)$ discussed 
here is, indeed, a part of a 
much larger algebraic structure. We will explore this structure in a future work  \cite{future} and  present just a few preliminary remarks here.

The main observation is that the near zone expansion considered by Starobinsky  \cite{1973JETP...37...28S} also exhibits a hidden $SL(2,\mathbb{R})$ symmetry.
The corresponding $SL(2,\mathbb{R})$ generators take the following form
\be
J_{a}=L_{a}+ {\Omega\beta}v_{0,a}\d_\phi,
\ee
where
\begin{gather}
v_{0,\pm 1}=e^{\pm\beta^{-1} t}\le{r-r_+\over r-r_-}\ri^{1/2}
\nonumber\\
\nonumber
v_{0,0}=-1\;.
\end{gather}
 This near zone expansion is less suited for demonstrating vanishing of static Love numbers at $\Omega m\neq 0$, 
 but appears to have other particularly nice properties.  For instance, arguments analogous to the ones presented above, prove
  that in the Starobinsky near zone approximation the black hole response vanishes at the locking frequency $\omega=m\Omega$. Furthermore, 
  explicit calculations demonstrate that in this approxmation all nonstatic Love numbers  vanish as well, which can also be proven algebraically~\cite{future}.

Note that by acting on $v_{0,a}$ with the $SL(2,\mathbb{R})$ generators (\ref{eq:vec0}) one obtains vectors
\begin{gather}
v_{0,n}=L_{-1}^{n-1}v_{0,1}=(-1)^{n-1}(n-1)!e^{-n\beta^{-1} t}\le{r-r_+\over r-r_-}\ri^{n/2}\nonumber\\
\nonumber
v_{0,-n}=L_{1}^{n-1}v_{0,-1}=(n-1)!e^{n\beta^{-1} t}\le{r-r_+\over r-r_-}\ri^{n/2}\;,
\end{gather}
where $n>0$.
Vectors $v_{0,k}$ with $k\in \mathbb Z$ are all regular at the past and future horizons and span an 
$SL(2,\mathbb{R})$ representation $V$ with zero Casimir, 
\[
{\cal C}_2(V)=0\;.
\]
These considerations suggest that it is natural to consider an infinite-dimensional extension of   the Love symmetry 
into a semidirect product $SL(2,\mathbb{R})\ltimes U(1)_V$, where $U(1)_V$  are vector fields of the form
$
v\d_\phi\;,
$
with $v\in V$. The near zone considered here and the one by Starobinsky correspond to different $SL(2,\mathbb{R})$ subalgebras of this larger algebra.

\section{Discussion and Future Directions}
The presented results open numerous new avenues for future research  both  on a purely theoretical side, and  as far as relations to gravitational wave observations are concerned. 
On a theory side, it is very satisfactory that the ``Love hierarchy problem" has led us to a novel symmetry. Static Love numbers vanish as a consequence of this symmetry.
 At first sight, everything is now consistent with the 't Hooft notion of naturalness  \cite{tHooft:1979rat}.

Note that the Love symmetry has an unconventional property that it mixes UV and IR modes. Indeed,
due to the presence of the $e^{\pm \beta^{-1} t}$ factors in $L_{\pm 1}$ generators, 
$SL(2,\mathbb{R})$ multiplets 
contain both the static solution
and high frequency 
modes.
However, only in the near extreme limit $\beta^{-1} M\ll1$ 
the action of the Love symmetry is compatible with the near zone conditions (\ref{eq:NZ}).
This does not invalidate any of our arguments. Our logic is first to take the near zone limit $\epsilon=0$, which provides accurate results for low frequency observables, and then to solve the resulting theory exactly. This allows us to benefit from the presence of the Love symmetry in spite of the UV/IR mixing introduced by  $L_{\pm 1}$
generators. Still, it is somewhat unclear whether this should be considered as a triumph of naturalness in the sense of 't Hooft, or rather an example of the ``UV miracle."
 It remains to be seen whether this unconventional example may provide useful lessons for other famous hierarchy problems.

It is  a popular slogan nowadays  that ``black holes are the hydrogen atom of 21st century", see, e.g., \cite{Hooft:2016cpw,hydrogen}. We  see that this comparison is actually accurate in a very concrete technical sense. Low energy dynamics of both systems is governed by an emergent integrable algebraic structure.
It is still natural to wonder who ordered these structures. What are the reasons for the $SO(4)$ Laplace--Runge--Lentz symmetry of the  hydrogen atom from the viewpoint of the full quantum electrodynamics and for the Love symmetry of black holes from the viewpoint of the full general relativity? 
We are not aware of a good answer in the hydrogen case, but it looks plausible that, for black holes, 
the horizon is the culprit. We already saw that nonzero static Love numbers for higher-dimensional black holes do not signal the loss of symmetry. It will be interesting to study what happens in 
other examples, such as in the presence of higher-derivative corrections to the Einstein action, c.f.~\cite{Cardoso:2018ptl}.

Other next natural steps in theoretical studies of the Love symmetry include a comprehensive analysis of its algebraic structure, understanding its relation to near horizon isometries in the extreme limit and  to the asymptotic Bondi--Metzner--Sachs symmetries,  and inclusion of massive fields. It will also be interesting to see whether unitary $SL(2,\mathbb{R})$ representations play any special role in this story.

At the same time, it is important to remember that the study of black hole responses is far from being a pure theorist's exercise. 
These effects contribute to gravitational waveforms of binary inspirals, and the corresponding Wilson coefficients will be probed by the forthcoming gravitational wave observations \cite{Porto:2016zng,Cardoso:2017cfl}. 
An approximate hidden symmetry provides an extremely valuable addition and a useful organizing principle to the effective field theory toolbox. 
Chiral symmetry of pion interactions is one of the most famous and successful illustrations of this. Similar to the pion case, it is important to systematically work out all consequences of the Love symmetry, including the ones beyond the strict static limit. 
To achieve this it should be fruitful to replace low frequency expansion with the near zone expansion. 
By treating the symmetry breaking parameters in (\ref{V1}) as spurions under the Love symmetry, 
it should be possible to obtain analogues of the Gell-Mann--Okubo relations for finite frequency responses and quasinormal modes.

\textit{Acknowledgments.}
We thank Mina
Arvanitaki,
Vitya Gorbenko, Lam Hui, 
and
Riccardo Rattazzi
for helpful discussions.
This work is supported in part by the NSF award PHY-1915219 and by the BSF grant 2018068.
MI is partially supported by the Simons Foundation’s Origins of the Universe Program.

\appendix 
\section{Conventions for the Kerr metric}
\label{App:met}
The Kerr metric in the Boyer--Lindquist coordinates takes the following form
\be
\label{Kerr}
\begin{split}
ds^2= &-\left(1-\frac{2Mr}{\Sigma}\right)dt^2
- \left(\frac{4 M a r \sin^2\theta}{\Sigma} \right)dtd\phi
+\frac{\Sigma}{\Delta}dr^2\\ +\Sigma d\theta^2
& +\sin^2\theta\left(r^2+a^2+\frac{2 M a^2r\sin^2\theta}{\Sigma}\right)d\phi^2\,,
\end{split}
\ee
where $M$ is the black hole mass, $0<a<M$ is the reduced spin parameter and
\[
\Delta=r^2-2Mr+a^2\;,\;\;\Sigma= r^2+a^2\cos^2 \theta\;.
\]
Two roots of $\Delta$ correspond to the outer $r_+$ and the inner $r_-$ black hole horizons. The horizon angular velocity is defined
as 
\[
\Omega={a\over 2Mr_+}\;.
\]
After the variable separation (\ref{sep}) the angular eigenfunctions satisfy~\cite{Teukolsky:1972my}
\be
\label{ang}
\begin{split}
\le-\frac{1}{\sin\theta}\d_\theta \sin\theta\d_\theta
 +\frac{m^2}{\sin^2\theta}+\epsilon a^2\omega^2\sin^2\theta  \ri S =\ell(\ell+1)S\,,
\end{split}
\ee
where $\epsilon$ is the near zone expansion parameter.

\section{Higher dimensions}
\label{sec:highd}
$SL(2,\mathbb{R})$ symmetry of the near zone Teukolsky equation for a  
 Schwarzschild $d$-dimensional black hole has been constructed in  \cite{Bertini:2011ga}. 
Its generators take the following form
\vspace{-0.1cm}
\be
\label{eq:vec0d}
\begin{split}
& L_0 = - \beta\d_t \,,\\
& L_{\pm} = e^{\pm\beta^{-1} t}\Bigg(
\mp \Delta^{1/2}_\r\d_\r + \beta\d_\r \Delta^{1/2}_\r\d_t
\Bigg)\,,
\end{split}
\ee
with $\r=r^{d-3}$, $\Delta_\r=\r(\r-r_+^{d-3})$ and
\be 
\beta=\frac{2r_+}{d-3}\,,
\quad 
r_+
=\frac{8\pi\Gamma((d-2)/2)M}{(d-2)\pi^{\frac{d-1}{2}}}\,.
\ee
The Klein-Gordon equation in the near zone 
can be written using the corresponding quadratic   
Casimir as 
\be
\mathcal{C}_2\Phi=
\left(\d_\r\Delta_\r \d_\r -\frac{r_+^{2d-4}}{(d-3)^2\Delta_\r}\d_t^2\right)\Phi
=\hat{\ell}(\hat{\ell}+1)\Phi\,,
\ee
where $\hat{\ell}\equiv \ell/(d-3)$.

\bibliography{short2.bib}

\end{document}